\documentclass[12pt,preprint]{aastex}
\begin{document}

\title{\bf Dependence of Heliospheric Ly$\alpha$ Absorption on the
  Interstellar Magnetic Field}

\author{Brian E. Wood\altaffilmark{1}, Vladislav V.
  Izmodenov\altaffilmark{2,3}, Jeffrey L. Linsky\altaffilmark{1},
  Dmitry Alexashov\altaffilmark{3} }

\altaffiltext{1}{JILA, University of Colorado, 440 UCB, Boulder, CO
  80309-0440; woodb@origins.colorado.edu, jlinsky@jila.colorado.edu.}
\altaffiltext{2}{Lomonosov Moscow State University, Dept. of
  Aeromechanics and Gas Dynamics, Moscow 119899, Russia; izmod@ipmnet.ru.}
\altaffiltext{3}{Institute for Problems in Mechanics RAS, Prospekt
  Vernadskogo 101-1, Moscow 117526, Russia; and Space Research
  Institute (IKI) RAS}

\begin{abstract}

     We use newly developed 3D kinetic MHD models of the heliosphere
to predict heliospheric H~I Ly$\alpha$ absorption for various lines
of sight.  These predictions are compared
with actual Ly$\alpha$ spectra from the {\em Hubble Space Telescope},
many of which have yielded previous detections of heliospheric
absorption.  We find that the absorption predicted by the models
is weakly affected by both the magnitude and orientation of the
assumed ISM magnetic field.  Models with $B=1.25-2.5$~$\mu$G and
an angle of $\alpha=15-45^{\circ}$ with respect to the upwind direction
of the ISM flow generally provide the best fits to the data,
but the sensitivity of the Ly$\alpha$ absorption to many model
input parameters makes it difficult to fully characterize the
region of parameter space allowed by the data.  We also use the models
to assess the degree to which heliospheric asymmetries induced by the
ISM field should be apparent in Ly$\alpha$ absorption.  An ISM field
that is skewed with respect to the ISM flow vector results in substantial
azimuthal asymmetries in both the hydrogen wall and heliosheath,
but only the heliosheath asymmetries yield potentially detectable
asymmetries in Ly$\alpha$ absorption; and then only in downwind
directions, where comparison with the data is complicated by few
actual absorption detections and an insufficient model grid extent.

\end{abstract}


\section{INTRODUCTION}

     The interaction region between the solar wind and ambient ISM has
been the subject of hydrodynamic modeling efforts \citep{enp61,enp63}
since around the time of the first {\it in situ} observations of the
solar wind by {\em Mariner~2} \citep{mn62} and by
{\em Luna~2} \citep{kig60}.  As shown in Figure~1, this interaction
results in a large scale structure for the heliosphere that consists of
three boundaries: the termination shock (TS), where the solar wind is
shocked to subsonic speeds; the bow shock (BS), where the ISM flow
is shocked to subsonic speeds; and in between the two the
heliopause (HP), which separates the plasma flows of the fully
ionized solar wind and partially ionized ISM. Reviews of the
history of heliospheric modeling include \citet{teh89}, \citet{vbb90},
\citet{gpz99}, and \citet{vbb06}.

     {\em Voyager~1} recently encountered the TS at a distance
of 94~AU from the Sun in roughly the upwind direction relative to the
ISM flow \citep{ecs05}.  However, the locations of the
more distant HP and BS remain observationally uncertain.
In general, there are very few observational constraints for the
properties of the heliosphere beyond the TS.  One of the
few exceptions is heliospheric Ly$\alpha$ absorption,
which is observable in {\em Hubble Space Telescope} (HST) spectra of
nearby stars.

     Unlike the ionized component of the ISM, the neutrals in the ISM
can penetrate into all regions of the heliosphere.  Charge exchange
processes involving these neutrals create populations of hot H~I
that permeate the heliosphere, and it is these neutrals that produce
absorption signatures in stellar Ly$\alpha$ lines observed by HST.
For most lines of sight, the absorption is dominated by H~I in the
so-called ``hydrogen wall'' region in between the HP and BS
\citep{vbb91,bew05b}, but in downwind
directions absorption from heliosheath neutrals, created by charge
exchange between the TS and HP, can be paramount
\citep{vvi99,bew07}.  The heliospheric absorption
is only detectable when the ISM absorption for the observed line of sight
is not too broad to obscure the absorption.  In upwind directions the
interstellar H~I column density (in cm$^{-2}$) must be
$\log N({\rm H~I})<18.2$ to detect heliospheric absorption, but in
downwind directions one must have $\log N({\rm H~I})<17.8$ \citep{bew05a}.
Astrospheric absorption from the wind-ISM interaction region surrounding
the observed star can also sometimes be detected.

     Starting with \citet{kgg97}, there have been many attempts
to use the Ly$\alpha$ absorption observations to test heliospheric
models.  The hydrodynamic models are generally quite successful in
reproducing the observed amount of absorption, especially in upwind
directions where the hydrogen wall accounts for most of it
\citep{kgg97,vvi99,vvi02,bew00}.  The Ly$\alpha$ absorption therefore
represents a convincing detection of the hydrogen wall, and a validation
of the models that predicted it even before it was detected by HST.

     However, the exact amount of absorption predicted by the models is
dependent on the parameters that are assumed for the Local
Interstellar Cloud (LIC) in which the Sun resides \citep{rl92}.  Some
aspects of the ambient ISM are known very well, such as the LIC flow
speed and direction \citep[e.g.,][]{mw04,em04}, but
others are not known as precisely.  Thus, there has been hope that
the Ly$\alpha$ absorption can help constrain certain properties of
the ISM.  \citet{vvi02}, for example, experimented with
numerous different models assuming different combinations of ISM
proton and H~I densities. The absorption predicted by the models
does vary with the input parameters, but the absorption diagnostic
seems to have only a modest sensitivity to most input parameters
of interest, making it difficult to simply define a range of
parameters that are consistent with the data.  The dependence of
the predicted absorption on the nature of the hydrodynamic code
used in the modeling is also a problem \citep{bew00,vvi02}.
The source of this difficulty lies in
the complexity of H~I velocity distributions within the
heliosphere, which are non-Maxwellian \citep[e.g.,][]{vi01}
and therefore can only be modeled with fully kinetic or
complex multi-fluid codes.

     All of the models that have been compared with the data in the
past have been 2-dimensional, axisymmetric models.  Recently,
3-dimensional MHD models have become available that are capable of
considering the effects of the ISM magnetic field on heliospheric
structure, while still maintaining a sufficiently sophisticated
treatment of the neutrals to properly consider them and the plasma
in a self-consistent manner \citep{vi05,vvi06,nvp06}.
Figure~1 presents results of
calculations made with a 3D kinetic MHD model of the heliosphere by
\citet{vi05}.  It shows that the inclusion of even a modest
ISM field can indeed affect the shape of the global heliosphere.
We will determine here whether this also has significant effects
on the Ly$\alpha$ absorption. The nature of the magnetic field in
the ISM immediately outside the heliosphere is poorly known, so we
will also assess the sensitivity of the Ly$\alpha$ absorption to
changes in the assumed ISM field strength and orientation.  In
doing so, we consider many more HST-observed lines of sight than
have been used in prior data-model comparisons.

\section{THE CHOSEN SAMPLE OF HST Ly$\alpha$ OBSERVATIONS}

     The amount of heliospheric Ly$\alpha$ absorption depends greatly
on the direction of the observed line of sight.  The greatest
spatial dependence is on the poloidal angle $\theta$ between the
line of sight and the upwind direction of the ISM flow.  Clearly
it is advantageous to consider many different lines of sight with
a wide variety of $\theta$'s in comparing the heliospheric
absorption predicted by models with the data.  Considering a
variety of directions is even more important when testing 3D MHD
models, which can yield heliospheric structures that are not
axisymmetric and therefore will have absorption predictions that
are dependent on the azimuthal angle as well as being dependent on
$\theta$ (see Fig.~1).  Past data-model comparisons considered no
more than six HST-observed lines of sight, which individually have
either provided real detections of heliospheric absorption or
merely upper limits \citep{bew00,vvi02}.
This is a rather small number of lines of sight even for testing
axisymmetric models, let alone the 3D ones. However, the number of
heliospheric absorption detections has recently increased
significantly \citep{bew05b,bew07}, so it is
well worthwhile to reassess the sample of available HST data to
select a larger sample of spectra to test the 3D kinetic MHD
models.

     \citet{bew05b} provide a complete list of HST-observed
Ly$\alpha$ spectra that are appropriate for our purposes, all of which
have been analyzed to measure ISM H~I column densities, to search for
evidence of heliospheric/astrospheric absorption, and to measure stellar
Ly$\alpha$ fluxes corrected for the contaminating ISM absorption.
Figure~2 is a sky map in ecliptic coordinates of lines of sight
with a stellar Ly$\alpha$ line observed by HST.  All these spectra have
sufficient spectral resolution to permit a reasonably precise search for
heliospheric absorption.  The boxes indicate the 11 lines of sight that
actually yield detections of heliospheric absorption.  All the other
lines of sight yield nondetections.

     Many of the detections are clustered around the upwind direction of
the ISM flow.  The advantageous nature of upwind lines of sight for
detecting heliospheric absorption is consistent with model predictions,
which suggest that the deceleration of H~I in the hydrogen wall relative
to the ISM flow should be largest in these directions.  This results in
a greater separation of the heliospheric absorption from that of the ISM,
thereby making it easier to detect heliospheric absorption in upwind
lines of sight \citep{bew05b}.

     There is also a cluster of three detections very close to the
downwind direction.  Initial analysis of these Ly$\alpha$ spectra did
{\em not} yield detections \citep{bew05b}.  However, we have found
that the stellar Ly$\alpha$ profiles reconstructed for
$\theta>160^{\circ}$ lines of sight are systematically blueshifted from
the stellar rest frames, indicating the presence of very broad, shallow
absorption on the red side of the Ly$\alpha$ profiles \citep{bew07}.
This is the exactly the sort of absorption signature one expects from
heliosheath neutrals (as opposed to hydrogen wall neutrals).  Since
very downwind lines of sight looking down the tail of the heliosphere
will have very long path lengths through the heliosheath, it is
in the most downwind lines of sight where one might expect to see this
broad absorption.
Thus, we now consider these three lines of sight to have detections
of heliospheric absorption, though the nature of these detections is
rather different from the others.

     Our goal is to select a sample of HST-observed lines of sight
from Figure~2 to use for comparing observed and predicted heliospheric
Ly$\alpha$ absorption.  Obviously we start by choosing the 11
detections, which actually provide quantitative measurements of the
absorption.  We add to these detections nine nondetections (diamonds
in Fig.~2) that at least provide upper limits for the amount of
absorption that might be present in those directions.  These
nondetections are chosen to sample parts of the sky not covered by
the detections.  Another major selection criterion is ISM H~I column
density.  Lines of sight with low ISM column densities are preferable
since they provide more restrictive upper limits on heliospheric
absorption.  Data quality (i.e., resolution, signal-to-noise) also
plays a role in choosing which nondetections to consider.

     Numbered symbols in Figure~2 indicate the final sample of 20
lines of sight to be used in our data-model comparisons.  The stellar
identifications of these lines of sight are indicated in Figures~3 and 4,
along with the Ly$\alpha$ spectra, which are displayed in order of
increasing $\theta$.  We focus only on the red side
of the Ly$\alpha$ absorption profile, where the heliospheric
absorption resides.  The fluxes are normalized to the intrinsic stellar
Ly$\alpha$ profile reconstructed in the original analysis of the data.
We refer the reader to \citet{bew05b} and references therein to
see the full Ly$\alpha$ spectra and descriptions of their analysis.
The dotted green lines in the figure show only the ISM absorption
based on these analyses.  For the heliospheric absorption detections,
there is excess absorption observed beyond that from the ISM.
Successful heliospheric models should predict the right
amount of excess absorption to fit the data for these lines of sight.
For the nondetections, the ISM absorption fits the data reasonably well.
In these cases, successful heliospheric models should predict essentially
no significant absorption beyond that from the ISM.

     The three $\theta>160^{\circ}$ detections (\#18--\#20 in Figs.~2--4)
are special cases, as mentioned above.  The original reconstructed stellar
Ly$\alpha$ profiles suggest no heliospheric absorption, but the
blueshifts of these profiles away from their stellar rest
frames implies that these profiles are inaccurate.  We can infer the amount
of heliospheric absorption in these directions by constructing a stellar
profile forced to be centered on the stellar radial velocity and
then seeing how much of the red wing of that profile must be absorbed to
yield the original profile.  The shaded regions in Figure~3 for these
three downwind lines of sight indicate this excess
absorption, where the uncertainties are estimated by allowing the
stellar radial velocity to be $\pm 3$ km~s$^{-1}$ from its measured
value.  This excess absorption cannot be extended to lower velocities
closer to the center of the Ly$\alpha$ line because near line center
stellar Ly$\alpha$ profiles cannot be assumed to be symmetric and
centered on the stellar rest frame.  Stellar Ly$\alpha$ profiles often
have self-reversals near line center, which are often asymmetric.  For
details about all of this, see \citet{bew07}.  The important thing
to note here is that for
these three very downwind lines of sight, the absorption predicted by
the models should {\em not} fit the data but should instead fall within
the shaded regions.

     Finally, the requirement that the intrinsic stellar Ly$\alpha$
profile be within $\pm 3$ km~s$^{-1}$ of the stellar rest frame allows
us to compute upper limits for the amount of broad heliosheath
absorption that can be present for lines of sight without detected
heliospheric absorption.  Thick dashed lines in Figure~3 show these
upper absorption limits, but only for downwind lines of sight
($\theta>110^{\circ}$) where the broad heliosheath absorption is
potentially prominent.  Absorption predictions from the models must lie
above these limits to be consistent with the data.  The dashed lines
cannot be extended to low velocities close to line center for the
same reason that the shaded regions of the $\theta>160^{\circ}$ lines
of sight are limited to the wings of the line (see above).

\section{THE INTERSTELLAR MAGNETIC FIELD'S EFFECTS ON Ly$\alpha$
  ABSORPTION}

     Figures 3 and 4 compare the HST Ly$\alpha$ data with the
heliospheric absorption predicted by 3D kinetic MHD models of
the heliospheric interface \citep{vi05,vvi06}, assuming
various directions and magnitudes for the ISM field.  The models used
here are of the type initially developed by \citet{vbb93,vbb95},
with a fully kinetic treatment of neutral hydrogen
within the heliosphere to provide the most precise computations of
the velocity distribution functions of the neutrals.  \citet{vi05}
expanded the 2D axisymmetric Baranov \& Malama code to
a fully 3D geometry, and also added the capability of including an
interstellar magnetic field in the model.  The code separates all
heliospheric H atoms into several populations:  1. original interstellar
atoms and other atoms originating outside of the bow shock, 2. secondary
interstellar atoms originating between the bow shock and heliopause,
3. atoms originating between the heliopause and termination shock, and
4. atoms originating in the supersonic solar wind.  We calculate number
densities, temperatures, and bulk velocities for these populations along
the lines of sight toward the observed stars by taking moments of the
velocity distributions.  The heliospheric absorption for each line of
sight is computed from these traces of density, temperature, and flow
velocity.  With this methodology, we are making the approximation
that the velocity distribution functions of the individual populations
are locally Maxwellian.

\subsection{Previous Constraints on the Local ISM Field}

     The nature of the interstellar magnetic field surrounding the
Sun is poorly known, though some observational constraints exist.
The global Galactic field has a magnitude of $1.6\pm 0.2$~$\mu$G and
is directed towards a Galactic longitude of $l=96\pm 4^{\circ}$, but
there is substantial local variability \citep{rjr89}, meaning
that the actual local field could be significantly higher or lower and
could be in a completely different direction.

     A $\sim 4^{\circ}$ discrepancy exists between the flow vectors of
interstellar He and H within the solar system, and the most promising
explanation for this is that the LIC's magnetic field is skewed with
respect to the ISM flow seen by the Sun, which can deflect the flow
of interstellar hydrogen atoms in the heliosphere \citep{rl05}.
Helium atoms are not affected in this manner, since their charge
exchange cross sections are much lower than hydrogen and they are
therefore effectively blind to the presence of the heliosphere.
If correct, this interpretation identifies a plane in which the ISM
field must lie, which happens to be inconsistent with the orientation of
the global Galactic field.  \citet{mo06} have argued that
an ISM field that is $\alpha=30^{\circ}-60^{\circ}$ from the apparent
flow direction can potentially explain {\em Voyager~1} and {\em 2}
observations of energetic particles flowing inwards from the TS.
The satellites have both seen these particles, but flowing in opposite
directions.  \citet{mo06} demonstrate that asymmetries in the heliospheric
structure induced by a skewed ISM magnetic field can potentially
cause this effect, thanks in part to the satellites' positions on opposite
sides of the ecliptic plane.

     A magnetic field much stronger than the global Galactic
field has been proposed to explain an apparent pressure imbalance
between the hot, ionized plasma that dominates the Local Bubble and the
warm, partially neutral clouds that lie within it.  The Local Bubble,
the cavity in which the Sun resides \citep{rl03}, is believed to account
for much of the soft X-ray background radiation \citep[see also][]{rks06}.
These X-rays seem to suggest thermal pressures of
$P/k \sim 15,000$ cm$^{-3}$~K \citep{sls98}.  In contrast, the LIC and
other warm clouds within the Local Bubble appear to have much lower
pressures of $P/k \sim 2280$ cm$^{-3}$~K \citep{ebj02,sr04}.
Such a large pressure imbalance within the local ISM seems unlikely.

     One way out of this dilemma is to propose that the LIC is supported
by a strong magnetic field of order 7~$\mu$G \citep{dpc03,vf04}.
This would seem to be disallowed by heliospheric models,
which imply that such large magnetic pressures would force the termination
shock well inside the 94~AU distance measured by {\em Voyager~1}
\citep{gg97,ecs05}.  The only way to avoid this problem
is for the field to be nearly parallel to the ISM flow.  However,
this would contradict the evidence mentioned above that the LIC field is
{\em not} parallel to to the LIC flow (i.e., the discrepant H and He
flow vectors within the solar system, and the opposite flow directions of
termination shock particles observed by {\em Voyager~1} and {\em 2}).
Thus, large fields are still difficult to reconcile with heliospheric
observations and models.  An alternative solution to the Local Bubble
pressure problem is that the pressures normally estimated from the soft
X-ray emission are too high for various reasons:  contamination from
heliospheric foreground emission \citep{tec00,rl04a,dk06},
contamination from X-ray emission from the walls of the
Local Bubble \citep{rl04b}, and improper assumption of collisional
ionization equilibrium \citep{db01}.

\subsection{Absorption Constraints on the ISM Field}

     Most of our models are computed assuming a modest magnetic field
of $B=2.5$~$\mu$G, which can compress the TS somewhat depending on
the field orientation, but not enough to be inconsistent with the
{\em Voyager~1} encounter distance of 94~AU (see Fig.~1).  We assume
that the magnetic field is oriented within the plane suggested by
\citet{rl05}.  Figure~3 shows the absorption predicted
by models with different field orientations within this plane,
for angles ranging from parallel to the ISM flow ($\alpha=0^{\circ}$)
to perpendicular to the ISM flow ($\alpha=90^{\circ}$).  Table~1
lists the ecliptic and Galactic coordinates that correspond to
these field directions.  Figure~4 shows the absorption predictions
for three models with $\alpha=45^{\circ}$,
but with different field strengths of 0, 1.25, and 2.5~$\mu$G.
The ISM hydrogen and proton densities assumed in these models are
$n_\infty(H~I)=0.18$ cm$^{-3}$ and $n_\infty(H^+)=0.06$ cm$^{-3}$, the
ISM flow speed is $V_\infty=26.4$ km~s$^{-1}$, and the temperature is
$T_\infty=6400$~K \citep{vi05}.

     Figure~3 shows that in upwind directions, where most of the
heliospheric detections lie, the absorption decreases
with increasing $\alpha$.  To better
illustrate this behavior, Figure~5 plots the absorbed Ly$\alpha$
flux predicted by the Figure~3 models versus $\alpha$.
No one model clearly
fits the data better than all the others, presumably due to the
absorption dependence on $\alpha$ being only a modest one. This
also may be indicative of the systematic uncertainties in the
estimation of the ISM absorption. The 61~Vir line of sight seems
particularly discrepant.  This is probably due to difficulties in
removing geocoronal emission blended with the red side of the
Ly$\alpha$ absorption line \citep[see Fig.~1 in][]{bew05b}, so
this line of sight should be regarded with caution.

     The $\alpha=60^{\circ}$ and $\alpha=90^{\circ}$ models underestimate
the absorption in all upwind directions (see Figs.~3 and 5), so perhaps
these models should be considered inconsistent with the data.  If one
ignores the problematic 61~Vir line of sight, the $\alpha=0^{\circ}$
model is a worse fit to the data than the $\alpha=15^{\circ}-45^{\circ}$
for all but the 36~Oph line of sight.  Considering the evidence
described above that the ISM field is skewed from the flow direction,
$\alpha=0^{\circ}$ seems unlikely anyway.  Thus, we consider
$\alpha=15^{\circ}-45^{\circ}$ to represent the most likely field
orientations for the local ISM, which overlaps the
$\alpha=30^{\circ}-60^{\circ}$ range quoted by \citet{mo06}.

     These conclusions are based on $B=2.5$~$\mu$G models, but
Figure~4 shows that assuming $B=1.25$~$\mu$G does not change the
absorption very much.  Thus, at least for these low-to-moderate
field strengths our conclusions concerning $\alpha$ are relatively
sound. However, Figure~4 shows that decreasing the field all the
way to zero does change the absorption, with the absorption being
somewhat higher in upwind directions.
The greater amount of absorption upwind is somewhat
surprising considering that the hydrogen wall is narrower for
$B=0$ (see Fig.~1).  However, the inclusion of even a modest ISM
field weakens the bow shock and lowers the H~I density in the
hydrogen wall, more than offsetting the broader width of the wall.

     It should be stated that definitive conclusions are difficult
to make at this point since the Ly$\alpha$ absorption
is at least somewhat sensitive to other input parameters, such
as the assumed ISM densities and temperature that also have observational
uncertainties \citep{vvi02}, though not nearly as large as
those involving the magnetic field.  A time-consuming
thorough exploration of parameter space would be necessary to fully
characterize the constraints on ISM properties provided by the
Ly$\alpha$ data.

     We have focused so far on comparing the models with data in
upwind directions, where absorption from the hydrogen wall is
dominant. Any conclusions drawn from downwind directions, where
heliosheath absorption is more prominent, will be far more
tentative.  The primary reason for this is that the grid used for
our current models only extends 500~AU from the Sun,
which is not nearly far enough to capture all of the heliosheath
absorption for $\theta\gtrsim 120^{\circ}$.
Thus, the absorption predictions shown in
Figures~3 and 4 for these directions will
underestimate the amount of absorption that the models would
really predict if the grid were extended further downwind.

     Many of the models seem to predict too much absorption downwind
even with the limited grid extent, particularly in the velocity
range of $80-120$ km~s$^{-1}$.  But there is another potential
difficulty with downwind absorption that concerns the treatment of
the plasma in the models.  Although a fully kinetic treatment is
applied to the neutrals, the plasma is assumed to be a single Maxwellian
fluid throughout the heliosphere.  However, this is clearly a poor
approximation, since pickup ions, for example, have non-Maxwellian
velocity distributions and are not thermalized with the solar wind
inside the termination shock \citep[e.g.,][]{gg04}.
\citet{ygm06} have replaced the simple single-fluid plasma
treatment in the 2D Baranov \& Malama code with a complex multi-component
representation of the plasma.  This more sophisticated plasma treatment
does not result in significantly different hydrogen wall absorption in
upwind directions, but we have found that is {\em does} result
in a significant reduction in heliosheath absorption which can
potentially alleviate the problems these models have in predicting
too much downwind absorption \citep{bew07}.  The most meaningful
comparison with the data in downwind directions would therefore
require that our 3D models also utilize such a multi-component
plasma treatment, as well as having grids that extend far enough
downwind to capture all the heliosheath absorption.  We leave such
computationally intensive modeling for a future paper.

     We note that asymmetries in the heliospheric structure induced by
the ISM field are evident in the downwind absorption predicted by
the models.  For example, there is significant model dependence in the
absorption towards DK~UMa ($\theta=112^{\circ}$), but no significant
model dependence towards $\tau$~Cet ($\theta=123^{\circ}$).  Since
$\theta$ is similar for these stars but their azimuthal angles quite
different (see Fig.~2), this difference in behavior must be due to
azimuthal variability, which can only be due to magnetic field
induced asymmetries (see \S3.3).

     Finally, we note that \citet{nvp06} have also developed
3D MHD heliospheric models that treat neutrals in a self-consistent
manner with the plasma.  This code uses a less sophisticated 2-fluid
treatment for the neutral H velocity distributions, but it includes the
effects of the interplanetary magnetic field on heliospheric structure
as well as the ISM field.  Absorption has been computed for a limited
number of these models, yielding results qualitatively similar to
those reported here, with absorption decreasing with both $\alpha$ and
the magnetic field strength \citep{bew06}.  Like \citet{vi05},
\citet{nvp06} demonstrate that the 3D MHD
models can potentially reproduce the shift between the H and He flows
observed within the solar system \citep{rl05}, though they
emphasize that the magnitude of the shift depends not only on the
strength and orientation of the magnetic field, but also on the ISM
neutral hydrogen density.

\subsection{Quantifying Expected Absorption Asymmetries}

     Figure~1 illustrates that even a modest ISM field can result
in a heliospheric structure that is significantly asymmetric,
consistent with other models that also predict asymmetries of this
sort \citep{rr98,nvp06}.
In the bottom half of Figure~1, the heliosheath (between the TS and HP)
is narrower than in the upper half, but the hydrogen wall (between the HP
and BS) is wider.  One might imagine that this would result in
corresponding Ly$\alpha$ absorption asymmetries.  In other words, the
Ly$\alpha$ absorption should be azimuthally dependent as well as $\theta$
dependent.

     Some evidence that the models do indeed predict azimuthally
dependent absorption in downwind directions is mentioned in \S3.2.
However, in order to properly quantify the degree of absorption asymmetry
expected based on the models, it is necessary to compare absorption
predictions for lines of sight with identical $\theta$ values but different
azimuthal angles ($\phi$).  This cannot be easily done with the set of
observed directions in Figures~3--4, which are scattered randomly
about the sky.

     Thus, in Figure~6 we show the heliospheric Ly$\alpha$
absorption predicted by the $B=2.5$~$\mu$G, $\alpha=45^{\circ}$
model for various $\phi$ angles, with $\theta$ fixed in the five
panels of the figure.  The azimuthal angle is defined such that
$\phi=0^{\circ}$ and $\phi=180^{\circ}$ are in the plane of the
ISM magnetic field.  This is the plane in which the
heliospheric structure is portrayed in Figure~1.  The $\phi=0^{\circ}$
direction would be associated with the upper half of Figure~1,
and $\phi=180^{\circ}$ would be associated with the lower half.

     In general, the hydrogen wall will be responsible
for the steep, saturated absorption edges of the absorption
profiles in Figure~6, which are particularly prominent in upwind
directions (e.g., located at 20--30 km~s$^{-1}$ in the
$\theta=30^{\circ}$ panel), while the heliosheath is responsible
for the broad, unsaturated absorption wings that extend to high
velocities, which become more prominent in downwind directions.
Very little $\phi$ dependence is apparent in upwind directions.
This is a bit surprising given the hydrogen wall asymmetries
apparent in Figure~1.  However, it turns out that azimuthal
density variations in the hydrogen wall offset the azimuthal width
dependence.  For example, although the hydrogen wall is narrower
for $\phi=0^{\circ}$ (corresponding to the upper half of Fig.~1)
than for $\phi=180^{\circ}$ (corresponding to the lower half of
Fig.~1), this is offset by higher hydrogen wall densities in the
$\phi=0^{\circ}$ direction, so integrated H~I column densities are
actually not very different.

     Figure~6 shows that for $\theta=60-90^{\circ}$, the particularly
narrow hydrogen wall at $\phi=0^{\circ}$ results in somewhat less
hydrogen wall absorption than other directions, but the difference is so
small it would be very difficult to detect in practice.  In contrast, the
heliosheath is thicker at $\phi=0^{\circ}$ and Figure~6 shows that
this does in fact lead to the broad heliosheath
absorption being significantly stronger in this direction than others,
though it is only in downwind directions ($\theta>90^{\circ}$) where
this azimuthal dependence becomes potentially detectable.  As discussed
in \S3.2, actual comparisons with the data are currently problematic
in these directions.  The asymmetries seen for the $\theta=90^{\circ}$
and $\theta=120^{\circ}$ surely extend to $\theta=150^{\circ}$ as well,
but Figure~6 does not show this due to the limited grid extent of the
models.  According to Figure~6, the
heliosheath absorption is at a minimum in directions normal to the plane
of the ISM field ($\phi=90^{\circ}$ and $\phi=270^{\circ}$), indicative
of magnetic compression of the heliosphere in those directions,
leading to shorter distances through the heliosheath for those lines
of sight.

\section{SUMMARY}

     We have compared H~I Ly$\alpha$ absorption profiles predicted
by 3D kinetic MHD models of the heliosphere with a large
selection of Ly$\alpha$ lines observed by HST, including many
lines of sight with detected heliospheric absorption.  The primary
purpose of this comparison is to assess the sensitivity of the
predicted absorption to changes in the ISM magnetic field
properties assumed in the model.  Our results are as follows:
\begin{description}
\item[1.] We find that the H~I Ly$\alpha$ absorption has a modest
  sensitivity to both the strength and orientation of the ISM
  magnetic field.  Focusing on upwind directions where most of the
  HST detections of heliospheric absorption reside, the models
  presented here with $B=1.25-2.5$~$\mu$G and $\alpha=15-45^{\circ}$
  appear to provide the best fits to the data, consistent with
  constraints from other sources \citep{gg97,vi05,rl05,mo06}.
\item[2.] However, since the Ly$\alpha$ absorption is sensitive to other
  model input parameters, such as the ISM H~I density, which have not
  been varied here, the region of parameter space that yields acceptable
  fits to the data will be complex.  It will be very difficult,
  perhaps impossible, for the Ly$\alpha$ absorption by itself to
  yield a unique set of model input parameters that fit the data.
  Nevertheless, the dependence of the absorption on many ISM
  parameters means that the absorption does provide one constraint
  on heliospheric models that is worthy of consideration in
  assessing how precisely the models reproduce reality.
\item[3.] The models show that an ISM field that is skewed with respect
  to the ISM flow vector results in substantial azimuthal asymmetries in
  the heliospheric hydrogen wall.  Surprisingly, these asymmetries do not
  result in significant asymmetries in Ly$\alpha$ absorption from the
  hydrogen wall, since densities within the wall vary in such a way as to
  cancel out the effects of the spatial asymmetries on hydrogen wall
  column densities.
\item[4.] The models also show that a skewed ISM field results in
  significant azimuthal asymmetries in the heliosheath, and unlike
  the hydrogen wall these asymmetries {\em do} yield significant azimuthal
  absorption dependence, at least in downwind directions where the
  heliosheath absorption is prominent.  These directions are clearly the
  best places to look for azimuthal dependences in Ly$\alpha$ absorption,
  but there are problems with doing this in practice.  One is simply
  that we have few downwind detections of heliospheric Ly$\alpha$
  absorption.  Another is that the heliosheath absorption that dominates
  in downwind directions should ideally be modeled using a complex
  multi-fluid plasma treatment.  And finally, the model grid must
  be extended a much longer distance downwind than present models to
  capture all the heliosheath absorption in these directions.  We hope
  to perform such computationally intensive modeling in the future.
\end{description}

\acknowledgments

This work was supported by
NASA grant NNG05GD69G to the University of Colorado.  V. I. was also
supported by RFBR grant 04-02-16559, the ``Dynastia'' Foundation, and
the ``Foundation in Support of Russian Science''.

\clearpage

\begin{deluxetable}{ccccc}
\tabletypesize{\normalsize}
\tablecaption{Model ISM Field Orientations}
\tablecolumns{5}
\tablewidth{0pt}
\tablehead{
  \colhead{$\alpha$ (deg)} & \multicolumn{2}{c}{Ecliptic Coord.} &
    \multicolumn{2}{c}{Galactic Coord.} \\
  \colhead{} & \colhead{l$_{e}$ (deg)} & \colhead{b$_{e}$ (deg)} &
    \colhead{l (deg)} & \colhead{b (deg)}}
\startdata
0  &  74.7 & $ -5.2$ & 183.3 & $-15.9$ \\
15 &  66.9 & $-18.1$ & 191.9 & $-28.6$ \\
30 &  57.8 & $-30.6$ & 202.7 & $-40.6$ \\
45 &  46.1 & $-42.4$ & 218.2 & $-51.2$ \\
60 &  29.5 & $-52.4$ & 241.0 & $-58.7$ \\
90 & 335.8 & $-59.6$ & 298.2 & $-56.0$ \\
\enddata
\end{deluxetable}

\clearpage

\begin{figure}[p]
\plotone{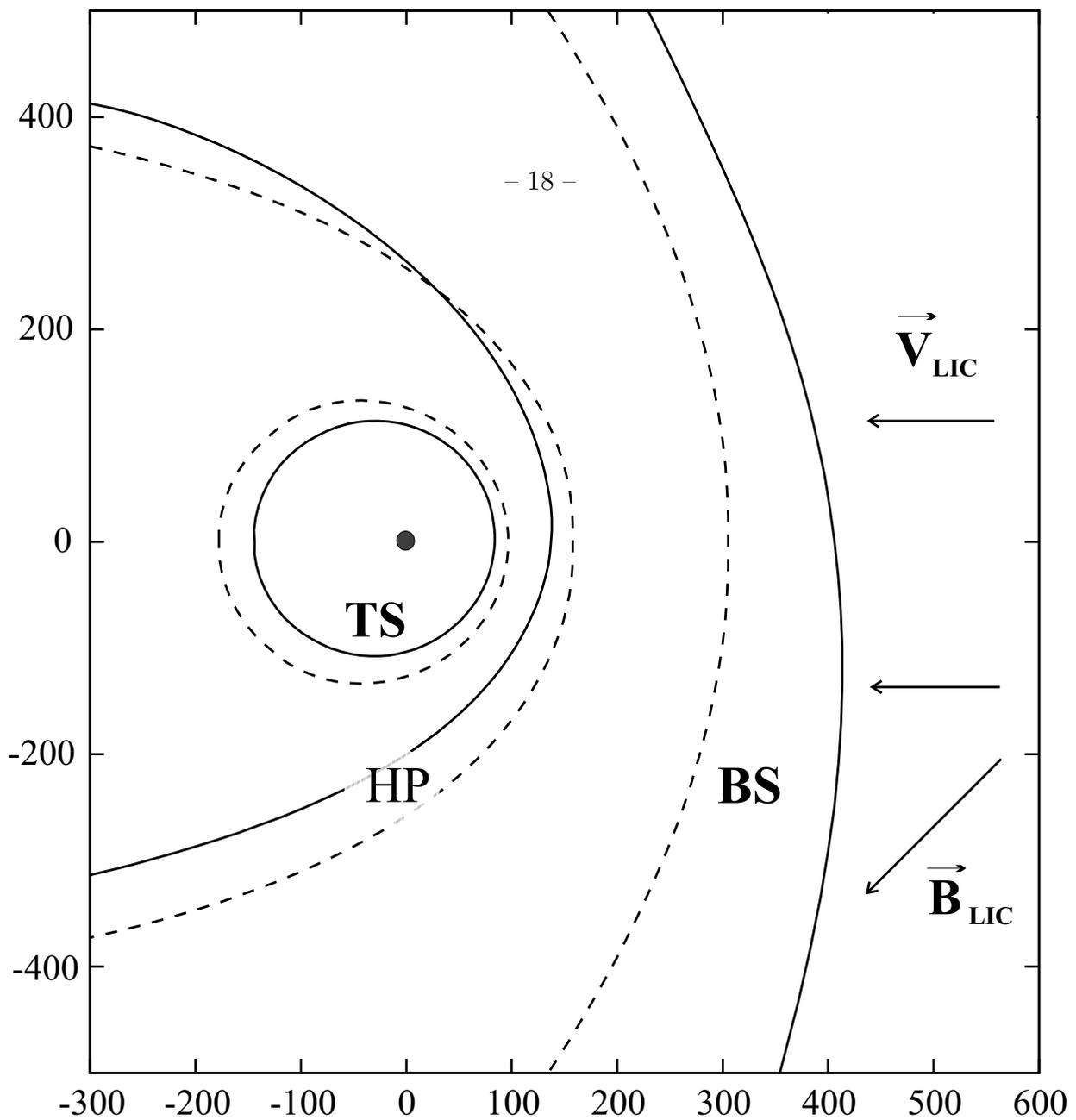}
\caption{The locations of the termination shock (TS), heliopause (HP),
  and bow shock (BS) for a model including a $B=2.5$ $\mu$G ISM field
  (solid lines), and a model with no ISM magnetic field (dashed lines).
  The directions of the LIC flow vector ($V_{LIC}$) and magnetic field
  ($B_{LIC}$) are indicated.  The distance scale is in AU.  The region
  between the HP and BS is sometimes called the ``hydrogen wall'' and
  the region between the TS and HP is the ``heliosheath.''}
\end{figure}

\begin{figure}[p]
\plotone{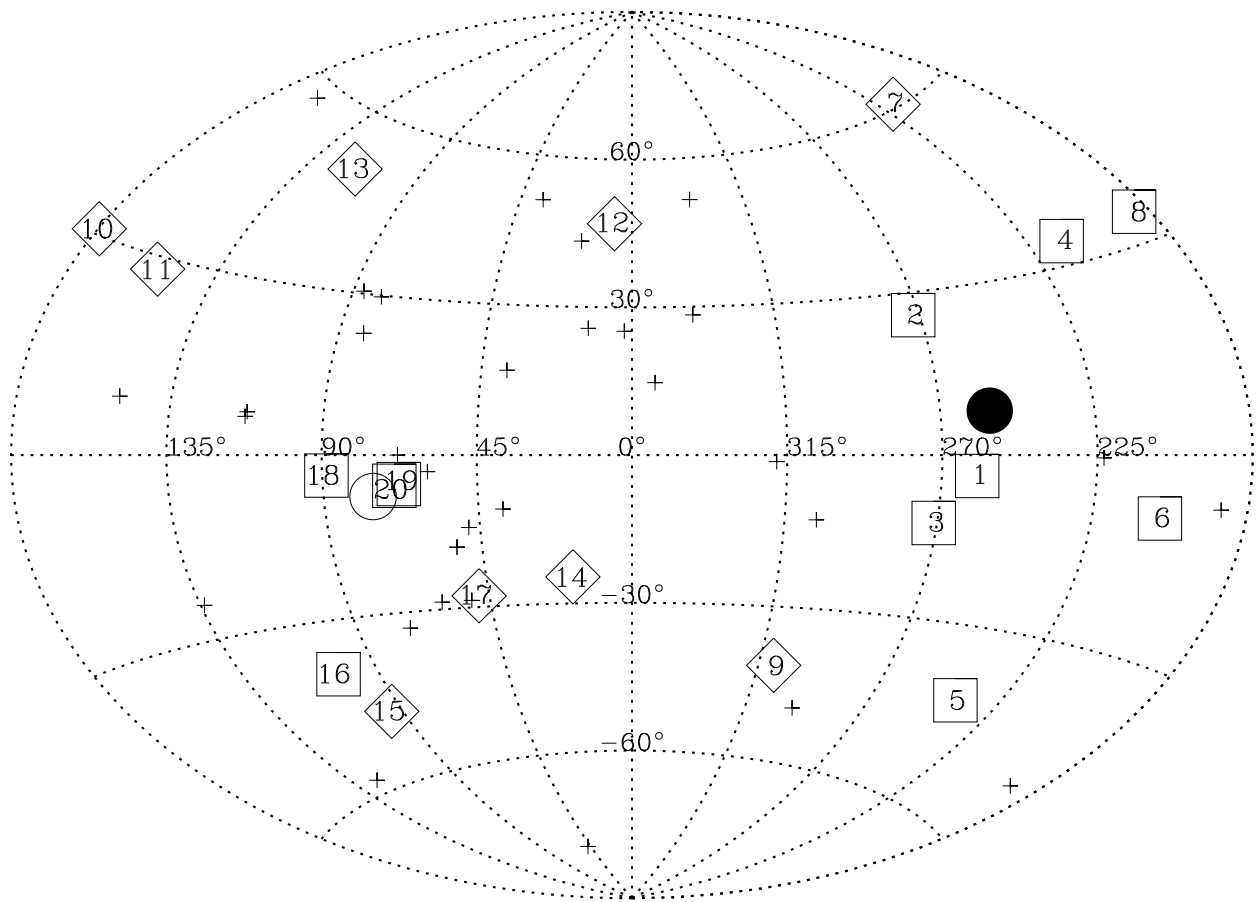}
\caption{Sky map in ecliptic coordinates of all HST-observed lines
  of sight with useful Ly$\alpha$ spectra.  The numbered symbols indicate
  spectra that we will compare with model predictions of Ly$\alpha$
  absorption (see Figs.~3 and 4).  Boxes indicate lines of
  sight with detected heliospheric absorption.  The plus signs and
  diamonds are both lines of sight with nondetections of heliospheric
  absorption.  The diamonds indicate lines of sight selected to
  provide upper limits for absorption in those directions.  The filled
  and open circles indicate the upwind and downwind directions of the
  local ISM flow vector, respectively.}
\end{figure}

\begin{figure}
\plotone{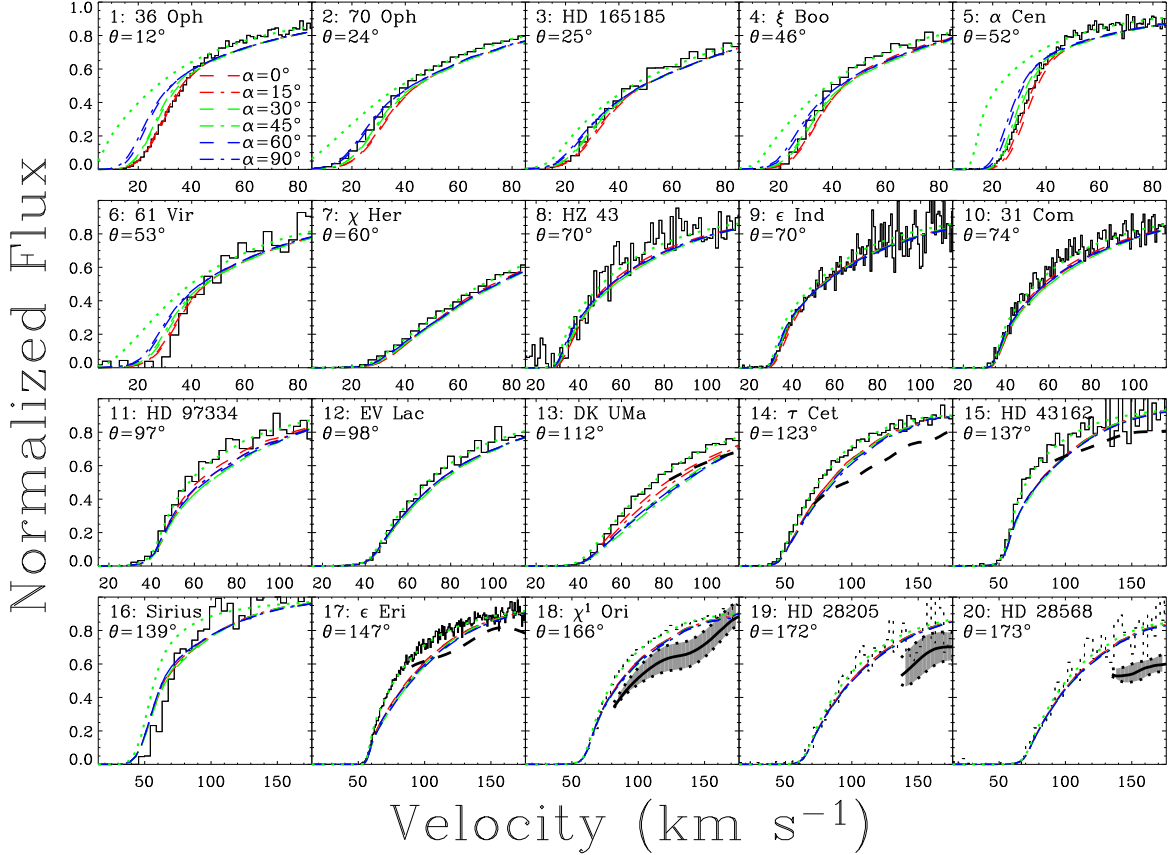}
\caption{The red side of the H~I Ly$\alpha$ absorption line (histogram)
  for the selected stars from Fig.~2, where the stars are placed in order
  of increasing angle from the upwind direction of the ISM flow ($\theta$).
  In each panel, the dotted green line is the ISM absorption alone.
  Absorption predictions are shown for heliospheric models computed
  assuming six different ISM field orientations, as quantified by
  $\alpha$, the angle between the field and the ISM flow direction
  (see the 36~Oph panel for line identifications).
  For many downwind lines of sight ($\theta > 110^{\circ}$),
  dashed lines show upper limits to the amount of absorption that can
  be present --- absorption predictions from the models must lie above
  these limits to be consistent with the data.  For the three most
  downwind lines of sight, the shaded regions indicate the amount of
  absorption that the models {\em should} predict if the real stellar
  Ly$\alpha$ profile is centered on the stellar rest frame rather
  than blueshifted as suggested by the original recontructed profile.
  For these lines of sight, the absorption predicted by the models
  should not fit the data (which are dotted histograms in these cases)
  but should instead fall within the shaded regions (see \S2).}
\end{figure}

\begin{figure}
\plotone{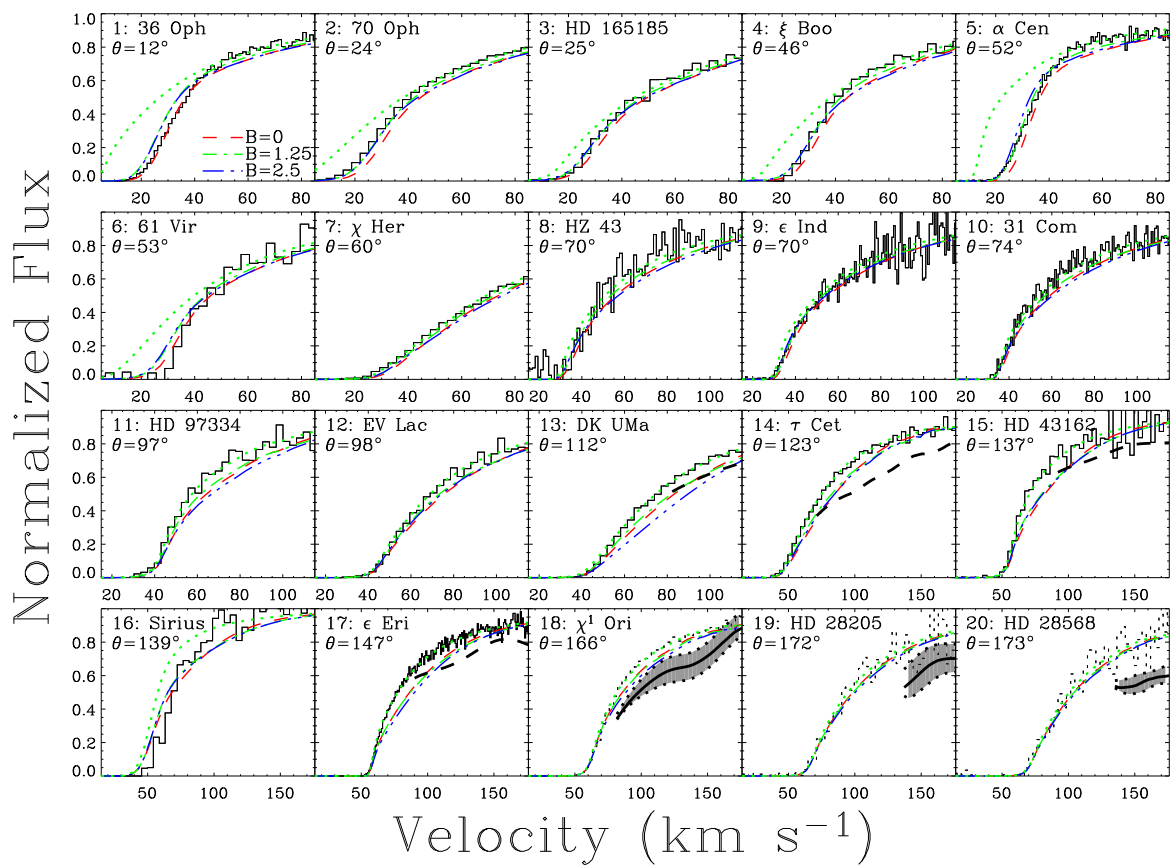}
\caption{A figure analogous to Fig.~3, but in this case the absorption
  predictions are for three $\alpha=45^{\circ}$ models that assume
  different ISM magnetic field strengths (see the 36~Oph panel for line
  identifications).}
\end{figure}

\begin{figure}
\plotone{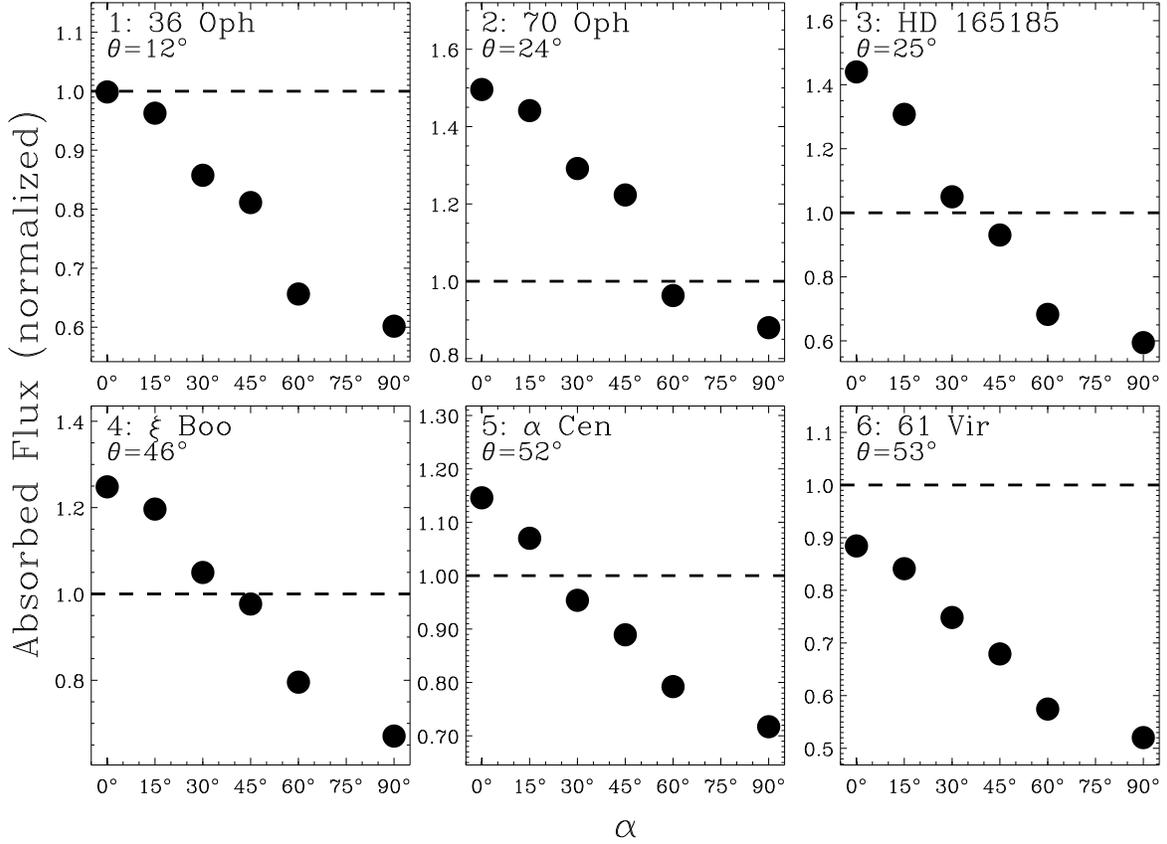}
\caption{For six upwind lines of sight, the predicted wavelength-integrated
  Ly$\alpha$ flux absorbed by heliospheric H~I beyond that absorbed by the
  ISM is computed for the six models from Fig.~3 and plotted versus
  $\alpha$, the ISM field orientation relative to the ISM flow direction.
  The fluxes are normalized to the observed amount of flux absorbed,
  so in each panel a flux of 1 (dashed lines) corresponds to agreement
  with the data.}
\end{figure}

\begin{figure}
\plotone{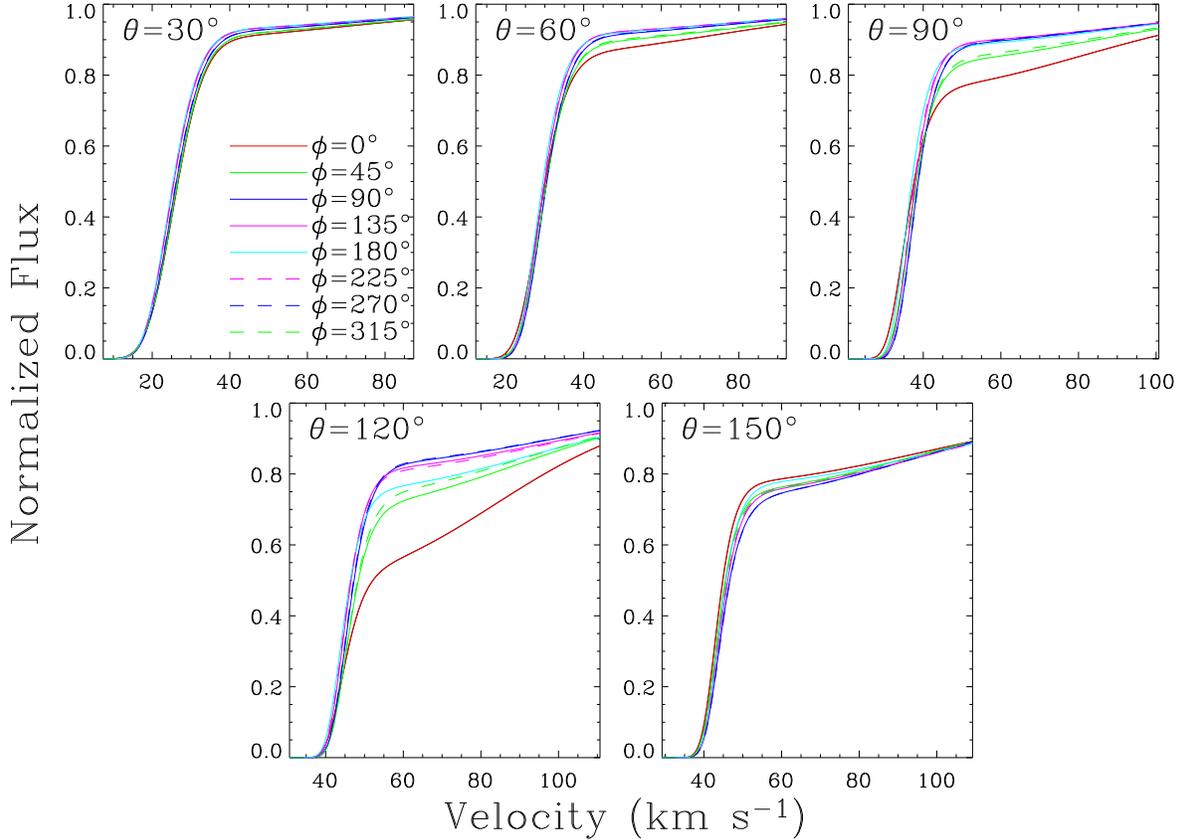}
\caption{An illustration of the directional dependence of
  H~I Ly$\alpha$ absorption predicted by a 3D MHD heliospheric model
  assuming an ISM field strength and orientation of $B=2.5$~$\mu$G and
  $\alpha=45^{\circ}$, respectively.  Absorption is shown for five
  values of the poloidal angle $\theta$ (the angle between the line
  of sight and the upwind direction of the ISM flow), and eight
  values of the azimuthal angle $\phi$ (where the plane of the ISM
  magnetic field is in the $\phi=0^{\circ}$ and $\phi=180^{\circ}$
  directions).  The model grid does not extend far enough downwind
  to properly search for azimuthal absorption variations at
  $\theta=150^{\circ}$.}
\end{figure}

\end{document}